# Single-state multi-party semiquantum key agreement protocol based on multi-particle GHZ entangled states

Tian-Jie Xu, Ying Chen, Mao-Jie Geng, Tian-Yu Ye*

College of Information & Electronic Engineering, Zhejiang Gongshang University, Hangzhou 310018, P.R.China

E-mail：yetianyu@mail.zjgsu.edu.cn

**Abstract:** In this paper, we put forward a novel single-state three-party semiquantum key agreement (SQKA) protocol with three-particle GHZ entangled states first. Different with previous quantum key agreement (QKA) protocols, the proposed single-state three-party SQKA protocol can realize the goal that a quantum party and two classical parties who only possess limited quantum capabilities equally contribute to the generation of a shared private key over quantum channels. Detailed security analysis turns out that the proposed single-state three-party SQKA protocol is secure against several famous attacks from an outside eavesdropper, such as the Trojan horse attack, the entangle-measure attack, the measure-resend attack and the intercept-resend attack. Moreover, it can resist the participant attack, which means that the shared private key cannot be determined fully by any nontrivial subset of three parties. The proposed single-state three-party SQKA protocol has the following nice features: (1) it only employs one kind of three-particle GHZ entangled states as initial quantum resource; (2) it doesn't need pre-shared keys among different parties; (3) it doesn't need unitary operations or quantum entanglement swapping. Finally, we generalize the proposed single-state three-party SQKA protocol into the case of $N$-party by only employing one kind of $N$-particle GHZ entangled states as initial quantum resource, which inherits the nice features of its three-party counterpart.

**Keywords:** Semiquantum cryptography; multi-party semiquantum key agreement; single state; GHZ entangled state

## 1  Introduction

In 1984, the first quantum key distribution (QKD) scheme was put forward by Bennett and Brassard [1], which means the appearance of quantum cryptography. Different from classical cryptography, whose security depends on the computing complexity of solving mathematical problems, quantum cryptography gains its theoretically unconditional security through the laws of physics such as quantum no-cloning theorem, Heisenberg's uncertainty principle *et al.* Since the birth of quantum cryptography, many people have been devoted to studying it [1-30]. As a result, various quantum cryptography branches have been established, such as QKD [1-6], quantum secure direct communication (QSDC) [7-14], quantum secret sharing (QSS) [15-19], quantum key agreement (QKA) [20-30] *et al.* QKA has been a hot topic of quantum cryptography in recent years. Different from QKD, in which one party distributes his private key to the other parties via quantum channels, QKA allows all participants to affect the shared key equally, which means that the shared key cannot be decided by any nontrivial subset of all participants.

In 2004, Zhou *et al.* [20] raised the first QKA protocol by utilizing quantum teleportation technique. However, Tsai and Hwang [21] subsequently pointed out a weakness in Zhou *et al.'s* protocol, i.e., the shared key can be fully determined by any participant. Later, Hsueh and Chen [22] proposed a QKA protocol based on maximally entangled states. In 2011, Chong *et al.* [23] proposed an improvement to two flaws of the protocol in Ref.[22]. In 2013, Shi and Zhong [24] presented the first multi-party QKA protocol using Bell states and entanglement swapping; Liu *et al.* [25] pointed out that the protocol in Ref.[24] cannot resist the participant attack, and then put forward a new multi-party QKA protocol with single particles; Yin *et al.* [26] presented a three-party QKA protocol with two-photon entanglement. In 2014, Xu *et al.* [27] suggested a novel multi-party QKA protocol with GHZ states. In 2016, Sun *et al.* [28] put forward an efficient multi-party QKA protocol with cluster states; Zhu *et al.* [29] pointed out that the protocol in Ref.[26] is not secure, as two dishonest participants can conspire to determine the shared key alone, and suggested an improvement to remedy this flaw. In 2020, Wang *et al.* [30] proposed a three-party QKA protocol with quantum Fourier transform.

The above QKA protocols always require all parties to possess full quantum capabilities，which may be impractical in some circumstances. In order to ease the burdens of quantum state preparation and quantum state measurement for partial parties, Boyer *et al.* [31-32] proposed the concept of semiquantum cryptography for the first time, which permits partial parties to only have limited quantum capabilities, such as measuring qubits with the $Z$-basis (i.e., $\{|0\rangle,|1\rangle\}$), preparing fresh qubits in the $Z$-basis, sending qubits without interference and reordering qubits through different delay lines. Since the birth of the concept of semiquantum cryptography, many scholars have applied it into different quantum cryptography branches to design different kinds of semiquantum cryptography protocols, such as semiquantum key distribution (SQKD) protocols [31-38], semiquantum private comparison (SQPC) protocols [39,40], semiquantum secret sharing (SQSS) protocols [41-43], semiquantum controlled secure direct communication (SQCSDC) protocol [44], semiquantum dialogue (SQD) protocol [44,45], semiquantum key agreement (SQKA) protocols [44,46,47], *et al*. With respect to SQKA, in 2017, Shukla *et al.* [44] proposed a two-party SQKA protocol with Bell states; Liu *et al.* [46] proposed a multi-party SQKA protocol with delegating quantum computation. In 2020, Zhou *et al.* [47] put forward a three-party SQKA protocol with four-particle cluster states.

Based on the above analysis, in this paper, in order to realize the goal that a quantum party and two classical parties equally contribute to the generation of a shared private key over quantum channels, we put forward a novel single-state three-party SQKA protocol by only adopting one kind of three-particle GHZ entangled states as initial quantum resource. Then, we generalize it into the case of $N$-party by using one kind of $N$-particle GHZ entangled states as initial quantum resource, which can realize the goal that a quantum party and $N-1$ classical parties equally contribute to the generation of a shared private key over quantum channels.

The remaining parts of this paper are arranged as follows: Section 2 describes the proposed single-state three-party SQKA protocol based on three-particle GHZ entangled states; Section 3 conducts its security analysis; Section 4 generalizes the proposed single-state three-party SQKA protocol into the one with $N$ parties; and finally, discussion and conclusion are given in Section 5.

## 2 The proposed single-state three-party SQKA protocol based on three-particle GHZ entangled states

Three-particle GHZ entangled states are three-qubit maximum entangled states, which are defined as

$$\left|\Psi^+_{0,0\oplus s,0\oplus t}\right\rangle_{ABC} = \frac{1}{\sqrt{2}}\left(\left|0,0\oplus s,0\oplus t\right\rangle + \left|1,1\oplus s,1\oplus t\right\rangle\right),$$

$$\left|\Psi^-_{0,0\oplus s,0\oplus t}\right\rangle_{ABC} = \frac{1}{\sqrt{2}}(-1)^{0\oplus t}\left(\left|0,0\oplus s,0\oplus t\right\rangle - \left|1,1\oplus s,1\oplus t\right\rangle\right), \quad (1)$$

where $s,t \in \{0,1\}$.

Suppose that Alice possesses unlimited quantum abilities while Bob and Charlie are two classical parties with limited quantum abilities. Alice, Bob and Charlie want to negotiate a shared private key over quantum channels on the condition that each participant equally contributes to the generation of this key which cannot be determined fully by any nontrivial subset of them. In order to accomplish this goal, we adopt the three-particle GHZ entangled state $\left|\Psi^+_{0,0,0}\right\rangle_{ABC}$ as initial quantum resource to design a single-state three-party SQKA protocol, which can be described as follows. For clarity, its flow chart is given in Fig.1. Note that throughout this protocol, the classical bit 0 corresponds to the state $\left|0\right\rangle$, while the classical bit 1 corresponds to the state $\left|1\right\rangle$.

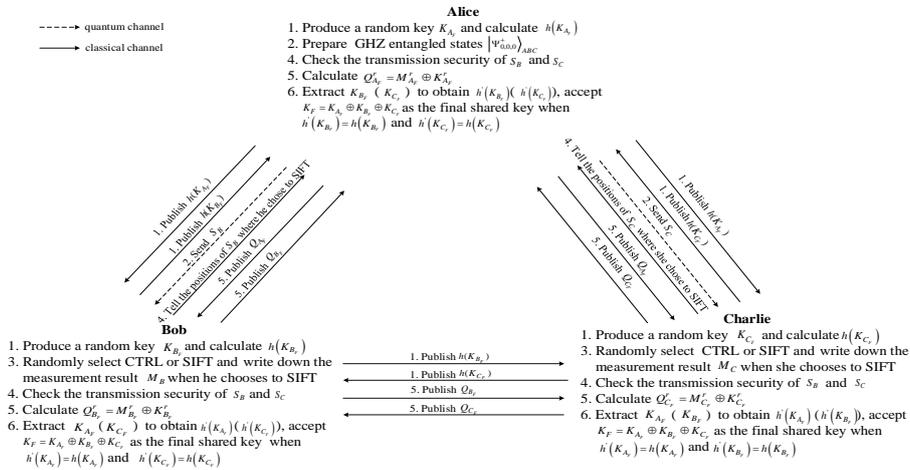

Fig. 1　The flow chart of the proposed single-state three-party SQKA protocol

Step 1: Alice, Bob and Charlie make use of quantum random number generator (QRNG) to produce the original random keys $K_{A_F}$, $K_{B_F}$ and $K_{C_F}$, respectively. Here, $K_{A_F} = \{K_{A_F}^1, K_{A_F}^2, ..., K_{A_F}^n\}$, $K_{B_F} = \{K_{B_F}^1, K_{B_F}^2, \cdots, K_{B_F}^n\}$, $K_{C_F} = \{K_{C_F}^1, K_{C_F}^2, \cdots, K_{C_F}^n\}$, where $K_{A_F}^r, K_{B_F}^r, K_{C_F}^r \in \{0,1\}$, $r = 1, 2, \cdots, n$ and $n$ is the length of shared private key. Each of Alice, Bob and Charlie calculates the hash value of the corresponding original random key and publishes the result to the other two parties. Here, $h(K_{A_F})$, $h(K_{B_F})$ and $h(K_{C_F})$ represents the hash values of $K_{A_F}$, $K_{B_F}$ and $K_{C_F}$, respectively.

Step 2: Alice prepares $8n$ GHZ entangled states all in the state of $|\Psi_{0,0,0}^+\rangle_{ABC}$. Then, Alice divides them into three sequences: $S_l = \{S_l^1, S_l^2, \cdots, S_l^{8n}\}$, where $S_l^j$ is the $j$ th particle of $S_l$, $l = A, B, C$ and $j = 1, 2, \cdots, 8n$. Specifically, all first, second and third particles of these GHZ entangled states form the ordered sequence $S_A$, $S_B$ and $S_C$, respectively. Alice sends the particles in $S_B$ ($S_C$) to Bob (Charlie) one by one. Note that after Alice sends the first particle to Bob (Charlie), she sends a particle only after receiving the previous one.

Step 3: For the $j$ th ($j = 1, 2, \cdots, 8n$) received particle in $S_B$ ($S_C$), Bob (Charlie) randomly executes one of the following two operations: ① the CTRL operation, i.e., reflecting the $j$ th received particle directly to Alice; and ② the SIFT operation, i.e., measuring the $j$ th received particle with the $Z$ basis to obtain the bit value of the measurement result $M_B^j$ ($M_C^j$), preparing a fresh particle in the same state as that he (she) found and resending it to Alice. Note that Bob (Charlie) chooses to SIFT and CTRL with equal probability; and there are only $4n$ particles Bob (Charlie) chooses to SIFT. Bob (Charlie) writes down $M_B = \{M_B^1, M_B^2, ..., M_B^{4n}\}$ ($M_C = \{M_C^1, M_C^2, ..., M_C^{4n}\}$) when he (she) chooses to SIFT.

Step 4: Bob (Charlie) tells Alice the positions of $S_B$ ($S_C$) where he (she) chose to SIFT. Alice performs different operations on the received particles according to Bob and Charlie's choices, as illustrated in Table 1. Case a, Case b and Case c are used to check whether the transmissions of $S_B$ and $S_C$ are secure or not; and Case d is used for not only checking the transmission security of $S_B$ and $S_C$ but also key agreement. Note that there are $2n$ positions where Case d happens; Alice randomly choose $n$ positions among these $2n$ ones to check the transmission security of $S_B$ and $S_C$.

In Case a, for the position where both Bob and Charlie chose to CTRL, Alice performs Action I. If there is no Eve on line, Alice's measurement result should be $|\Psi_{0,0,0}^+\rangle_{ABC}$;

In Case b, for the position where Bob chose to CTRL and Charlie chose to SIFT, Alice performs Action II. If there is no Eve on line, Charlie's measurement result on the original particle,

Alice's measurement result on the particle from Bob and Alice's measurement result on the corresponding particle in her own hand should be same; For checking whether there is an Eve or not, Charlie needs to tell Alice her measurement result;

In Case c, for the position where Bob chose to SIFT and Charlie chose to CTRL, Alice performs Action III. If there is no Eve on line, Bob's measurement result on the original particle, Alice's measurement result on the particle from Charlie and Alice's measurement result on the corresponding particle in her own hand should be same; For checking whether there is an Eve or not, Bob needs to tell Alice his measurement result;

In Case d, for the position where Bob chose to SIFT and Charlie chose to SIFT, Alice performs Action IV. If there is no Eve on line, Bob's measurement result on the original particle, Charlie's measurement result on the original particle and Alice's measurement result on the corresponding particle in her own hand should always be same; and Alice's measurement result on the particle from Bob (Charlie) should be same to the fresh state generated by Bob (Charlie). For checking whether there is an Eve or not, Bob and Charlie need to tell Alice their measurement results for the $n$ chosen positions;

If either of the error rates of these four Cases is abnormally high, the communication will be terminated immediately; otherwise, the communication will be continued.

Table 1    Alice's actions corresponding to those of Bob and Charlie

| Case | Bob | Charlie | Alice |
|------|------|---------|-----------|
| a | CTRL | CTRL | Action I |
| b | CTRL | SIFT | Action II |
| c | SIFT | CTRL | Action III |
| d | SIFT | SIFT | Action IV |

Action I: Alice measures the particle from Bob, the particle from Charlie together with the corresponding particle in her own hand with the GHZ basis;

Action II: Alice measures the particle from Bob and the corresponding particle in her own hand with the $Z$ basis;

Action III: Alice measures the particle from Charlie and the corresponding particle in her own hand with the $Z$ basis.

Action IV: Alice measures the particle from Bob, the particle from Charlie and the corresponding particle in her own hand with the $Z$ basis.

Step 5: As for Case d, after the $n$ particles used for security check are discarded, the remaining $n$ ones are used for key agreement, which are called as INFO particles for simplicity. The bit values of Alice's measurement results on the INFO particles in her own hand are represented by $M_{A_F} = \{M_{A_F}^1, M_{A_F}^2, ..., M_{A_F}^n\}$. Similarly, $M_{B_F} = \{M_{B_F}^1, M_{B_F}^2, ..., M_{B_F}^n\}$ denote the bits of $M_B$ corresponding to INFO particles, while $M_{C_F} = \{M_{C_F}^1, M_{C_F}^2, ..., M_{C_F}^n\}$ represent the bits of $M_C$ corresponding to INFO particles. It is apparent that

$$M_{A_F} = M_{B_F} = M_{C_F}. \tag{2}$$

Alice calculates

$$Q_{A_F}^r = M_{A_F}^r \oplus K_{A_F}^r, \tag{3}$$

where $r = 1, 2, \cdots, n$, and publishes $Q_{A_F}$ to Bob and Charlie, where $Q_{A_F} = \{Q_{A_F}^1, Q_{A_F}^2, ..., Q_{A_F}^n\}$. In the meanwhile, Bob calculates

$$Q_{B_F}^r = M_{B_F}^r \oplus K_{B_F}^r, \tag{4}$$

and publishes $Q_{B_F}$ to Charlie and Alice, where $Q_{B_F} = \{Q_{B_F}^1, Q_{B_F}^2, ..., Q_{B_F}^n\}$; Charlie calculates

$$Q_{C_F}^r = M_{C_F}^r \oplus K_{C_F}^r, \tag{5}$$

and publishes $Q_{C_F}$ to Alice and Bob, where $Q_{C_F} = \{Q_{C_F}^1, Q_{C_F}^2, ..., Q_{C_F}^n\}$.

Step 6: According to $M_{A_F}$, Alice extracts $K_{B_F}$ ($K_{C_F}$) from $Q_{B_F}$ ($Q_{C_F}$). Then, Alice calculates the hash value of $K_{B_F}$ ($K_{C_F}$) to obtain the result $h'(K_{B_F})$ ($h'(K_{C_F})$). If $h'(K_{B_F}) = h(K_{B_F})$ and $h'(K_{C_F}) = h(K_{C_F})$, Alice will accept $K_F = K_{A_F} \oplus K_{B_F} \oplus K_{C_F}$ as the final shared key.

Bob deduces $K_{A_F}$ ($K_{C_F}$) from $Q_{A_F}$ ($Q_{C_F}$) and $M_{B_F}$. Then, Bob calculates the hash value of $K_{A_F}$ ($K_{C_F}$) to obtain the result $h'(K_{A_F})$ ($h'(K_{C_F})$). If $h'(K_{A_F}) = h(K_{A_F})$ and $h'(K_{C_F}) = h(K_{C_F})$, Bob will accept $K_F = K_{A_F} \oplus K_{B_F} \oplus K_{C_F}$ as the final shared key.

Charlie deduces $K_{A_F}$ ($K_{B_F}$) from $Q_{A_F}$ ($Q_{B_F}$) and $M_{C_F}$. Then, Charlie calculates the hash value of $K_{A_F}$ ($K_{B_F}$) to obtain the result $h'(K_{A_F})$ ($h'(K_{B_F})$). If $h'(K_{A_F}) = h(K_{A_F})$ and $h'(K_{B_F}) = h(K_{B_F})$, Charlie will accept $K_F = K_{A_F} \oplus K_{B_F} \oplus K_{C_F}$ as the final shared key.

If any of Alice, Bob and Charlie refuses $K_F = K_{A_F} \oplus K_{B_F} \oplus K_{C_F}$ as the final shared key, the protocol will be terminated and restarted from the beginning.

## 3 Security analysis

### 3.1 Outside attack

In the proposed protocol, in order to obtain $K_F$, an outside eavesdropper, Eve may try her best to get something useful during the whole transmission processes of the particles of $S_B$ and $S_C$ by launching some famous attacks, such as the trojan horse attack, the entangle-measure attack, the measure-resend attack and the intercept-resend attack.

(1) The Trojan horse attack

In the proposed protocol, the particles of $S_B$ ($S_C$) go a round trip between Alice and Bob (Charlie), so we should take actions to overcome the Trojan horse attack from Eve, including the invisible photon eavesdropping attack [48] and the delay-photon Trojan horse attack [49,50]. As pointed out in Refs.[50,51], for defeating the invisible photon eavesdropping attack, Bob (Charlie) can put a wavelength filter in front of his (her) device to erase the illegitimate photon signal; and for preventing the delay-photon Trojan horse attack, Bob (Charlie) can adopt a photon number splitter (PNS: 50/50) to divide each sample signal into two parts, use the correct measuring bases to measure them and check whether the multiphoton rate is abnormally high or not.

(2) The entangle-measure attack

Eve's entangle-measure attack on the particles of $S_B$ and $S_C$ can be modeled as two unitaries: $U_E$ attacking particles from Alice to Bob and Charlie and $U_F$ attacking particles back from Bob and Charlie to Alice, where $U_E$ and $U_F$ share a common probe space with initial state $|\xi\rangle_E$. The shared probe allows Eve to launch her attack on the returned particles depending on the knowledge acquired from $U_E$ (if Eve does not make use of this fact, the "shared probe" can be regarded as the composite system formed by two independent probes) [31,32]. Any attack where Eve would make $U_F$ depend on a measurement after performing $U_E$ can be realized by $U_E$ and $U_F$ with controlled gates. Eve's entangle-measure attack on the particles of $S_B$ and $S_C$ can be depicted as Fig.2.

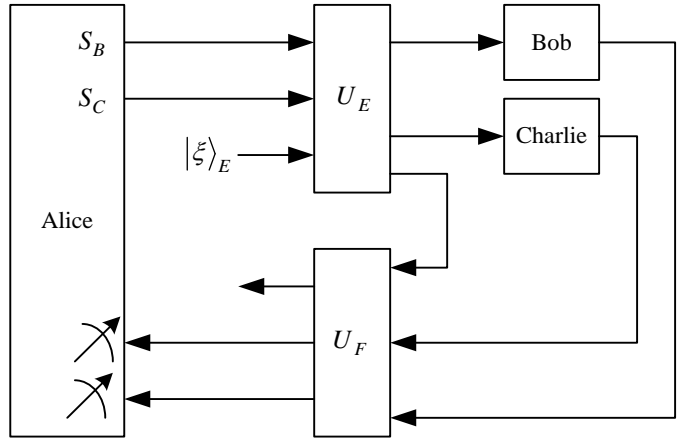

Fig.2 Eve's entangle-measure attack on the particles of $S_B$ and $S_C$ with two unitaries $U_E$ and $U_F$

**Theorem 1.** *Suppose that Eve performs attack $(U_E, U_F)$ on the particles from Alice to Bob and Charlie and back to Alice. For this attack inducing no error in Step 4, the final state of Eve's* *probe should be independent of Alice, Bob and Charlie's operations and their measurement results.*

**Proof.** The effect of $U_E$ on the qubits $|0\rangle$ and $|1\rangle$ can be described as

$$U_E(|0\rangle|\xi\rangle_E) = \beta_{00}|0\rangle|\xi_{00}\rangle + \beta_{01}|1\rangle|\xi_{01}\rangle, \tag{6}$$

$$U_E(|1\rangle|\xi\rangle_E) = \beta_{10}|0\rangle|\xi_{10}\rangle + \beta_{11}|1\rangle|\xi_{11}\rangle, \tag{7}$$

where $|\xi_{00}\rangle, |\xi_{01}\rangle, |\xi_{10}\rangle$ and $|\xi_{11}\rangle$ are Eve's probe states depending on $U_E$, $|\beta_{00}|^2 + |\beta_{01}|^2 = 1$ and $|\beta_{10}|^2 + |\beta_{11}|^2 = 1$.

According to Stinespring dilation theorem, the global state of the composite system before Bob and Charlie's operations is

$$\begin{aligned}
U_E\left(\left|\Psi_{0,0,0}^+\right\rangle_{ABC}|\xi\rangle_E\right) &= U_E\left[\frac{1}{\sqrt{2}}(|000\rangle + |111\rangle)_{ABC}|\xi\rangle_E\right] \\
&= \frac{1}{\sqrt{2}}\Big[|0\rangle_A\left(\beta_{00}|0\rangle_B|\xi_{00}\rangle + \beta_{01}|1\rangle_B|\xi_{01}\rangle\right)\left(\beta_{00}|0\rangle_C|\xi_{00}\rangle + \beta_{01}|1\rangle_C|\xi_{01}\rangle\right) \\
&\quad + |1\rangle_A\left(\beta_{10}|0\rangle_B|\xi_{10}\rangle + \beta_{11}|1\rangle_B|\xi_{11}\rangle\right)\left(\beta_{10}|0\rangle_C|\xi_{10}\rangle + \beta_{11}|1\rangle_C|\xi_{11}\rangle\right)\Big] \\
&= \frac{1}{\sqrt{2}}\Big[|0\rangle_A|0\rangle_B|0\rangle_C\left(\beta_{00}^2|\xi_{00}\rangle|\xi_{00}\rangle\right) + |0\rangle_A|0\rangle_B|1\rangle_C\left(\beta_{00}\beta_{01}|\xi_{00}\rangle|\xi_{01}\rangle\right) \\
&\quad + |0\rangle_A|1\rangle_B|0\rangle_C\left(\beta_{01}\beta_{00}|\xi_{01}\rangle|\xi_{00}\rangle\right) + |0\rangle_A|1\rangle_B|1\rangle_C\left(\beta_{01}^2|\xi_{01}\rangle|\xi_{01}\rangle\right) \\
&\quad + |1\rangle_A|0\rangle_B|0\rangle_C\left(\beta_{10}^2|\xi_{10}\rangle|\xi_{10}\rangle\right) + |1\rangle_A|0\rangle_B|1\rangle_C\left(\beta_{10}\beta_{11}|\xi_{10}\rangle|\xi_{11}\rangle\right) \\
&\quad + |1\rangle_A|1\rangle_B|0\rangle_C\left(\beta_{11}\beta_{10}|\xi_{11}\rangle|\xi_{10}\rangle\right) + |1\rangle_A|1\rangle_B|1\rangle_C\left(\beta_{11}^2|\xi_{11}\rangle|\xi_{11}\rangle\right)\Big], \\
&= |0\rangle_A|0\rangle_B|0\rangle_C|E_{000}\rangle + |0\rangle_A|0\rangle_B|1\rangle_C|E_{001}\rangle + |0\rangle_A|1\rangle_B|0\rangle_C|E_{010}\rangle \\
&\quad + |0\rangle_A|1\rangle_B|1\rangle_C|E_{011}\rangle + |1\rangle_A|0\rangle_B|0\rangle_C|E_{100}\rangle + |1\rangle_A|0\rangle_B|1\rangle_C|E_{101}\rangle \\
&\quad + |1\rangle_A|1\rangle_B|0\rangle_C|E_{110}\rangle + |1\rangle_A|1\rangle_B|1\rangle_C|E_{111}\rangle, \tag{8}
\end{aligned}$$

where the subscripts $A$, $B$ and $C$ denote the particles from $S_A$, $S_B$ and $S_C$, respectively, and $|E_{000}\rangle = \beta_{00}^2|\xi_{00}\rangle|\xi_{00}\rangle$, $|E_{001}\rangle = \beta_{00}\beta_{01}|\xi_{00}\rangle|\xi_{01}\rangle$, $|E_{010}\rangle = \beta_{01}\beta_{00}|\xi_{01}\rangle|\xi_{00}\rangle$, $|E_{011}\rangle = \beta_{01}^2|\xi_{01}\rangle|\xi_{01}\rangle$, $|E_{100}\rangle = \beta_{10}^2|\xi_{10}\rangle|\xi_{10}\rangle$, $|E_{101}\rangle = \beta_{10}\beta_{11}|\xi_{10}\rangle|\xi_{11}\rangle$, $|E_{110}\rangle = \beta_{11}\beta_{10}|\xi_{11}\rangle|\xi_{10}\rangle$, $|E_{111}\rangle = \beta_{11}^2|\xi_{11}\rangle|\xi_{11}\rangle$.

When Bob and Charlie receive the particles from Alice, they choose either to SIFT or to CTRL. After that, Eve performs $U_F$ on the particles sent back to Alice.

(i) Firstly, consider the case that both Bob and Charlie choose to SIFT. As a result, the state of

the composite system of Eq.(8) is collapsed into $|x\rangle_A |y\rangle_B |z\rangle_C |E_{xyz}\rangle$, where $x, y, z \in \{0,1\}$. For Eve not being detectable in Step 4, Bob's measurement result on the original particle, Charlie's measurement result on the original particle and Alice's measurement result on the corresponding particle in her own hand should always be same. Hence, we have

$$|E_{001}\rangle = |E_{010}\rangle = |E_{011}\rangle = |E_{100}\rangle = |E_{101}\rangle = |E_{110}\rangle = 0. \tag{9}$$

Moreover, Alice's measurement result on the particle from Bob (Charlie) should be same to the fresh state generated by Bob (Charlie), thus $U_F$ should satisfy

$$U_F \left( |x\rangle_A |y\rangle_B |z\rangle_C |E_{xyz}\rangle \right) = |x\rangle_A |y\rangle_B |z\rangle_C |F_{xyz}\rangle, \quad x = y = z \in \{0,1\}, \tag{10}$$

which means that $U_F$ cannot alter the states of the particles from Alice, Bob and Charlie after Bob and Charlie's SIFT operations. Otherwise, Eve is discovered with a non-zero probability.

(ii) Secondly, consider the case that Bob chooses to SIFT and Charlie chooses to CTRL. As a result, if Bob's measurement result is $|0\rangle_B$, the state of the composite system of Eq.(8) will be collapsed into $|0\rangle_A |0\rangle_B |0\rangle_C |E_{000}\rangle + |0\rangle_A |0\rangle_B |1\rangle_C |E_{001}\rangle + |1\rangle_A |0\rangle_B |0\rangle_C |E_{100}\rangle + |1\rangle_A |0\rangle_B |1\rangle_C |E_{101}\rangle$; and if Bob's measurement result is $|1\rangle_B$, the state of the composite system of Eq.(8) will be collapsed into $|0\rangle_A |1\rangle_B |0\rangle_C |E_{010}\rangle + |0\rangle_A |1\rangle_B |1\rangle_C |E_{011}\rangle + |1\rangle_A |1\rangle_B |0\rangle_C |E_{110}\rangle + |1\rangle_A |1\rangle_B |1\rangle_C |E_{111}\rangle$.

Suppose that Bob's measurement result is $|0\rangle_B$. After Eve performs $U_F$ on the particles sent from Bob and Charlie back to Alice, due to Eq.(9) and Eq.(10), the state of the composite system is evolved into

$$U_F \left( |0\rangle_A |0\rangle_B |0\rangle_C |E_{000}\rangle + |0\rangle_A |0\rangle_B |1\rangle_C |E_{001}\rangle + |1\rangle_A |0\rangle_B |0\rangle_C |E_{100}\rangle + |1\rangle_A |0\rangle_B |1_c\rangle_C |E_{101}\rangle \right)$$

$$= |0\rangle_A |0\rangle_B |0\rangle_C |F_{000}\rangle. \tag{11}$$

For Eve not being detectable in Step 4, Alice's measurement results on her own corresponding particle and the particle reflected by Charlie should be in the states of $|0\rangle_A$ and $|0\rangle_C$, respectively. Apparently, Eq.(11) naturally satisfies this requirement.

On the other hand, assume that Bob's measurement result is $|1\rangle_B$. After Eve performs $U_F$ on the particles sent from Bob and Charlie back to Alice, due to Eq.(9) and Eq.(10), the state of the composite system is evolved into

$$U_F \left( |0\rangle_A |1\rangle_B |0\rangle_C |E_{010}\rangle + |0\rangle_A |1\rangle_B |1\rangle_C |E_{011}\rangle + |1\rangle_A |1\rangle_B |0\rangle_C |E_{110}\rangle + |1\rangle_A |1\rangle_B |1\rangle_C |E_{111}\rangle \right)$$

$$= |1\rangle_A |1\rangle_B |1\rangle_C |F_{111}\rangle. \tag{12}$$

For Eve not being detectable in Step 4, Alice's measurement results on her own corresponding particle and the particle reflected by Charlie should be in the states of $|1\rangle_A$ and $|1\rangle_C$, respectively. Apparently, Eq.(12) naturally satisfies this requirement.

(iii) Thirdly, consider the case that Bob chooses to CTRL and Charlie chooses to SIFT. As a result, if Charlie's measurement result is $|0\rangle_C$, the state of the composite system of Eq.(8) will be collapsed into $|0\rangle_A|0\rangle_B|0\rangle_C|E_{000}\rangle + |0\rangle_A|1\rangle_B|0\rangle_C|E_{010}\rangle + |1\rangle_A|0\rangle_B|0\rangle_C|E_{100}\rangle + |1\rangle_A|1\rangle_B|0\rangle_C|E_{110}\rangle$;

and if Charlie's measurement result is $|1\rangle_C$, the state of the composite system of Eq.(8) will be collapsed into $|0\rangle_A|0\rangle_B|1\rangle_C|E_{001}\rangle + |0\rangle_A|1\rangle_B|1\rangle_C|E_{011}\rangle + |1\rangle_A|0\rangle_B|1\rangle_C|E_{101}\rangle + |1\rangle_A|1\rangle_B|1\rangle_C|E_{111}\rangle$.

Suppose that Charlie's measurement result is $|0\rangle_C$. After Eve performs $U_F$ on the particles sent from Bob and Charlie back to Alice, due to Eq.(9) and Eq.(10), the state of the composite system is evolved into

$$U_F\left(|0\rangle_A|0\rangle_B|0\rangle_C|E_{000}\rangle + |0\rangle_A|1\rangle_B|0\rangle_C|E_{010}\rangle + |1\rangle_A|0\rangle_B|0\rangle_C|E_{100}\rangle + |1\rangle_A|1\rangle_B|0\rangle_C|E_{110}\rangle\right)$$
$$= |0\rangle_A|0\rangle_B|0\rangle_C|F_{000}\rangle. \qquad (13)$$

For Eve not being detectable in Step 4, Alice's measurement results on her own corresponding particle and the particle reflected by Bob should be in the states of $|0\rangle_A$ and $|0\rangle_B$, respectively. Apparently, Eq.(13) naturally satisfies this requirement.

On the other hand, assume that Charlie's measurement result is $|1\rangle_C$. After Eve performs $U_F$ on the particles sent from Bob and Charlie back to Alice, due to Eq.(9) and Eq.(10), the state of the composite system is evolved into

$$U_F\left(|0\rangle_A|0\rangle_B|1\rangle_C|E_{001}\rangle + |0\rangle_A|1\rangle_B|1\rangle_C|E_{011}\rangle + |1\rangle_A|0\rangle_B|1\rangle_C|E_{101}\rangle + |1\rangle_A|1\rangle_B|1\rangle_C|E_{111}\rangle\right)$$
$$= |1\rangle_A|1\rangle_B|1\rangle_C|F_{111}\rangle. \qquad (14)$$

For Eve not being detectable in Step 4, Alice's measurement results on her own corresponding particle and the particle reflected by Bob should be in the states of $|1\rangle_A$ and $|1\rangle_B$, respectively. Apparently, Eq.(14) naturally satisfies this requirement.

(iv) Fourthly, consider the case that both Bob and Charlie choose to CTRL. After Eve performs $U_F$ on the particles sent back to Alice, due to Eq.(9) and Eq.(10), the state of the composite system is evolved into

$$U_F\left(|0\rangle_A|0\rangle_B|0\rangle_C|E_{000}\rangle + |0\rangle_A|0\rangle_B|1\rangle_C|E_{001}\rangle + |0\rangle_A|1\rangle_B|0\rangle_C|E_{010}\rangle + |0\rangle_A|1\rangle_B|1\rangle_C|E_{011}\rangle\right.$$
$$\left. + |1\rangle_A|0\rangle_B|0\rangle_C|E_{100}\rangle + |1\rangle_A|0\rangle_B|1\rangle_C|E_{101}\rangle + |1\rangle_A|1\rangle_B|0\rangle_C|E_{110}\rangle + |1\rangle_A|1\rangle_B|1\rangle_C|E_{111}\rangle\right)$$

$$= |0\rangle_A |0\rangle_B |0\rangle_C |F_{000}\rangle + |1\rangle_A |1\rangle_B |1\rangle_C |F_{111}\rangle$$

$$= \frac{1}{\sqrt{2}} \left( \left|\Psi^+_{0,0,0}\right\rangle_{ABC} + \left|\Psi^-_{0,0,0}\right\rangle_{ABC} \right) |F_{000}\rangle + \frac{1}{\sqrt{2}} \left( \left|\Psi^+_{0,0,0}\right\rangle_{ABC} - \left|\Psi^-_{0,0,0}\right\rangle_{ABC} \right) |F_{111}\rangle$$

$$= \frac{1}{\sqrt{2}} \left|\Psi^+_{0,0,0}\right\rangle_{ABC} (|F_{000}\rangle + |F_{111}\rangle) + \frac{1}{\sqrt{2}} \left|\Psi^-_{0,0,0}\right\rangle_{ABC} (|F_{000}\rangle - |F_{111}\rangle). \quad (15)$$

For Eve not being detectable in Step 4, Alice's measurement results on the particle reflected by Bob, the corresponding particle reflected by Charlie and her own corresponding particle should be in the state of $\left|\Psi^+_{0,0,0}\right\rangle_{ABC}$. Thus, it can be obtained from Eq.(15) that

$$|F_{000}\rangle = |F_{111}\rangle = |F\rangle. \quad (16)$$

(v) Inserting Eq.(16) into Eq.(10) derives

$$U_F \left( |x\rangle_A |y\rangle_B |z\rangle_C |E_{xyz}\rangle \right) = |x\rangle_A |y\rangle_B |z\rangle_C |F\rangle, \quad x = y = z \in \{0,1\}. \quad (17)$$

Inserting Eq.(16) into Eq.(11) generates

$$U_F \left( |0\rangle_A |0\rangle_B |0\rangle_C |E_{000}\rangle + |0\rangle_A |0\rangle_B |1\rangle_C |E_{001}\rangle + |1\rangle_A |0\rangle_B |0\rangle_C |E_{100}\rangle + |1\rangle_A |0\rangle_B |1_c\rangle_C |E_{101}\rangle \right)$$

$$= |0\rangle_A |0\rangle_B |0\rangle_C |F\rangle. \quad (18)$$

Inserting Eq.(16) into Eq.(12) produces

$$U_F \left( |0\rangle_A |1\rangle_B |0\rangle_C |E_{010}\rangle + |0\rangle_A |1\rangle_B |1\rangle_C |E_{011}\rangle + |1\rangle_A |1\rangle_B |0\rangle_C |E_{110}\rangle + |1\rangle_A |1\rangle_B |1\rangle_C |E_{111}\rangle \right)$$

$$= |1\rangle_A |1\rangle_B |1\rangle_C |F\rangle. \quad (19)$$

Inserting Eq.(16) into Eq.(13) derives

$$U_F \left( |0\rangle_A |0\rangle_B |0\rangle_C |E_{000}\rangle + |0\rangle_A |1\rangle_B |0\rangle_C |E_{010}\rangle + |1\rangle_A |0\rangle_B |0\rangle_C |E_{100}\rangle + |1\rangle_A |1\rangle_B |0\rangle_C |E_{110}\rangle \right)$$

$$= |0\rangle_A |0\rangle_B |0\rangle_C |F\rangle. \quad (20)$$

Inserting Eq.(16) into Eq.(14) generates

$$U_F \left( |0\rangle_A |0\rangle_B |1\rangle_C |E_{001}\rangle + |0\rangle_A |1\rangle_B |1\rangle_C |E_{011}\rangle + |1\rangle_A |0\rangle_B |1\rangle_C |E_{101}\rangle + |1\rangle_A |1\rangle_B |1\rangle_C |E_{111}\rangle \right)$$

$$= |1\rangle_A |1\rangle_B |1\rangle_C |F\rangle. \quad (21)$$

Inserting Eq.(16) into Eq.(15) produces

$$U_F \big( |0\rangle_A |0\rangle_B |0\rangle_C |E_{000}\rangle + |0\rangle_A |0\rangle_B |1\rangle_C |E_{001}\rangle + |0\rangle_A |1\rangle_B |0\rangle_C |E_{010}\rangle + |0\rangle_A |1\rangle_B |1\rangle_C |E_{011}\rangle$$

$$+ |1\rangle_A |0\rangle_B |0\rangle_C |E_{100}\rangle + |1\rangle_A |0\rangle_B |1\rangle_C |E_{101}\rangle + |1\rangle_A |1\rangle_B |0\rangle_C |E_{110}\rangle + |1\rangle_A |1\rangle_B |1\rangle_C |E_{111}\rangle \big)$$

$$= \sqrt{2} \left|\Psi^+_{0,0,0}\right\rangle_{ABC} |F\rangle. \quad (22)$$

It can be concluded from Eqs.(17-22) that for this attack inducing no error in Step 4, the final state of Eve's probe should be independent of Alice, Bob and Charlie's operations and their measurement results.

It should be pointed out that when Alice publishes $Q_{A_F}$ to Bob and Charlie, Eve may hear $Q_{A_F}$. However, Eve still cannot decode out $K_{A_F}$ from $Q_{A_F}$, as she has no idea about $M_{A_F}$. Likewise, when Bob (Charlie) publishes $Q_{B_F}$ ($Q_{C_F}$) to Charlie (Bob) and Alice, Eve may hear $Q_{B_F}$ ($Q_{C_F}$). However, Eve still cannot decode out $K_{B_F}$ ($K_{C_F}$) from $Q_{B_F}$ ($Q_{C_F}$), as she has no idea about $M_{B_F}$ ($M_{C_F}$). As a result, Eve cannot get $K_F$ by launching this kind of entangle-measure attack.

(3) The measure-resend attack

In order to know $M_B$ and $M_C$, Eve may intercept the particles of $S_B$ and $S_C$ sent from Alice, measure them with the $Z$ basis and resend the resulted states back to Bob and Charlie. After Eve's measurement, the initial state prepared by Alice is collapsed into $|000\rangle_{ABC}$ or $|111\rangle_{ABC}$ with the same probability. Without loss of generality, suppose that the $j$ th initial state prepared by Alice is collapsed into $|000\rangle_{ABC}$ after Eve's measurement. When both Bob and Charlie choose to CTRL, Bob and Charlie reflect the received particles back to Alice; and Alice measures the particle from Bob, the particle from Charlie and the corresponding particle in her hand with the GHZ basis, and obtains $|\Psi^+_{0,0,0}\rangle_{ABC}$ or $|\Psi^-_{0,0,0}\rangle_{ABC}$ with the same probability. Hence, Eve is detected by Alice, Bob and Charlie with the probability of $\frac{1}{2}$ in this case. When Bob chooses to CTRL and Charlie chooses to SIFT, Bob reflects the received particle back to Alice; Charlie measures the received particle with the $Z$ basis to obtain the measurement result $|0\rangle_C$, prepares a fresh particle in the same state as that she found and sends it back to Alice; and Alice measures the particle from Bob and the corresponding particle in her own hand with the $Z$ basis. Hence, Eve is detected by Alice, Bob and Charlie with the probability of 0 in this case. Likewise, when Bob chooses to SIFT and Charlie chooses to CTRL or both Bob and Charlie choose to SIFT, Eve is also detected by Alice, Bob and Charlie with the probability of 0. To sum up, when Eve launches this kind of measure-resend attack, she is detected with the probability of $\frac{1}{4} \times \frac{1}{2} + \frac{1}{4} \times 0 + \frac{1}{4} \times 0 + \frac{1}{4} \times 0 = \frac{1}{8}$. Therefore, Alice, Bob and Charlie have a probability of $1 - \left(\frac{7}{8}\right)^{7n}$ to detect Eve, which will converge to 1 if $n$ is large enough. The reason why Eve's measure-resend attack can be detected lies in two aspects: on one hand, the entanglement correlation among different particles of the initial state is destroyed by Eve's measurement; on the other hand, Bob and Charlie's operations are random to Eve.

(4) The intercept-resend attack

In order to know $M_B$ ($M_C$), Eve may prepare the fake sequence $S_B^*$ ($S_C^*$) in the $Z$ basis beforehand, intercept the particles of $S_B$ ($S_C$) sent from Alice, and send the particles of $S_B^*$ ($S_C^*$) to Bob (Charlie). Without loss of generality, take the $j$th fake particle in $S_B^*$ ($S_C^*$) being $|0\rangle_{E_B}$ ($|0\rangle_{E_C}$) for example here. When both Bob and Charlie choose to CTRL, Bob and Charlie reflect their respective $j$th received fake particles to Alice; and Alice measures the particle from Bob, the particle from Charlie and the corresponding particle in her hand with the GHZ basis, and obtains $|\Psi_{0,0,0}^+\rangle_{AE_BE_C}$, $|\Psi_{0,0,0}^-\rangle_{AE_BE_C}$, $|\Psi_{0,1,1}^+\rangle_{AE_BE_C}$ or $|\Psi_{0,1,1}^-\rangle_{AE_BE_C}$ with the same probability. Hence, Eve is detected by Alice, Bob and Charlie with the probability of $\frac{3}{4}$ in this case. When Bob chooses to CTRL and Charlie chooses to SIFT, Bob reflects the $j$th received fake particle back to Alice; Charlie measures the $j$th received fake particle with the $Z$ basis to obtain the measurement result $|0\rangle_{E_C}$, prepares a fresh particle in the same state as that she found and sends it back to Alice; and Alice measures the particle from Bob and the corresponding particle in her own hand with the $Z$ basis. Hence, Eve is detected by Alice, Bob and Charlie with the probability of $\frac{1}{2}$ in this case. Likewise, when Bob chooses to SIFT and Charlie chooses to CTRL or both Bob and Charlie choose to SIFT, Eve is also detected by Alice, Bob and Charlie with the probability of $\frac{1}{2}$. To sum up, when Eve launches this kind of intercept-resend attack, she is detected with the probability of $\frac{1}{4}\times\frac{3}{4}+\frac{1}{4}\times\frac{1}{2}+\frac{1}{4}\times\frac{1}{2}+\frac{1}{4}\times\frac{1}{2}\times\frac{1}{2}=\frac{1}{2}$. Therefore, the probability that Alice, Bob and Charlie can detect Eve is $1-\left(\frac{1}{2}\right)^{7n}$, which will converge to 1 if $n$ is large enough. Eve's intercept-resend attack can be discovered for the two reasons: on one hand, Eve's fake particles are likely to be different from the genuine ones; on the other hand, Bob and Charlie's operations are unpredictable to Eve.

3.2 Inside attack

In 2013, Sun *et al.* [52] pointed out that a QKA protocol should have the fairness property, i.e., all involved participants equally contribute to the final shared private key. In other words, non-trivial subset of the participants cannot determine the final shared private key. In the following, we will validate that the proposed three-party SQKA protocol can satisfy this requirement.

In the proposed three-party SQKA protocol, as long as all of Alice, Bob and Charlie honestly implement the protocol before Step 5, they can obtain $M_{A_F} = M_{B_F} = M_{C_F}$, which are used to encrypt their respective random keys later. If they are clever enough, it will be not necessary for them to launch attacks before Step 5 of the protocol. In Step 5, Alice publishes $Q_{A_F}$ to Bob and Charlie, Bob publishes $Q_{B_F}$ to Charlie and Alice, and Charlie publishes $Q_{C_F}$ to Alice and Bob. Hence, they may take the chance of publishing their encrypted results to conduct the cheating behavior. Without loss of generality, suppose that Alice is dishonest and wants to set the final shared key to be $K^*$ alone. After hearing of $Q_{B_F}$ and $Q_{C_F}$, she can decode out $K_{B_F}$ and $K_{C_F}$, respectively, according to $M_{A_F}$. Then, Alice calculates

$$Q_{A_F}^{*r} = M_{A_F}^r \oplus \left( K^{*r} \oplus K_{B_F}^r \oplus K_{C_F}^r \right), \tag{23}$$

where $K^{*r}$ is the $r$ th bit of $K^*$, and $r = 1, 2, \cdots, n$. Afterward, Alice publishes $Q_{A_F}^*$ to Bob and Charlie, where $Q_{A_F}^* = \left\{ Q_{A_F}^{*1}, Q_{A_F}^{*2}, ..., Q_{A_F}^{*n} \right\}$. After hearing of $Q_{A_F}^*$ and $Q_{C_F}$ ($Q_{B_F}$), according to $M_{B_F}$ ($M_{C_F}$), Bob (Charlie) can decode out $K^* \oplus K_{B_F} \oplus K_{C_F}$ and $K_{C_F}$ ($K_{B_F}$), respectively. As a result, Bob (Charlie) obtains the fake final key $K^*$ by calculating $\left( K^* \oplus K_{B_F} \oplus K_{C_F} \right) \oplus K_{B_F} \oplus K_{C_F} = K^*$. It looks like that Alice's cheating behavior is successful. However, in Step 6, Bob (Charlie) calculates the hash value of $K^* \oplus K_{B_F} \oplus K_{C_F}$ and discovers $h^{'}\left( K^* \oplus K_{B_F} \oplus K_{C_F} \right) \neq h\left( K_{A_F} \right)$. As a result, Bob (Charlie) refuses $K^*$ as the final shared key. Hence, Alice's cheating behavior fails.

If two dishonest parties collude together to conduct the cheating behavior similar to the above one, due to the usage of hash function, their cheating behavior will be also inevitably discovered by the third party.

It can be concluded now that the proposed three-party SQKA protocol can resist the inside attack and possess the fairness property, due to the usage of the collision resistance property of hash function [44].

## 4 The proposed single-state $N$-party SQKA protocol based on $N$-particle GHZ entangled states

Suppose that there are $N$ ($N > 3$) parties $P_1, P_2, \cdots, P_N$, among which $P_1$ possesses unlimited quantum abilities and $P_2, P_3, \cdots, P_N$ only have limited quantum abilities. $P_1, P_2, \cdots, P_N$ want to negotiate a shared private key over quantum channels together on the condition that each party equally contributes to the production of this key, which cannot be fully decided by any nontrivial subset of them. In order to achieve this aim, we generalize the above single-state three-party SQKA protocol into the one with $N$ parties by adopting the $N$-particle GHZ entangled state $\left|\Psi^+_{0,0,\cdots,0}\right\rangle_N = \frac{1}{\sqrt{2}}\left(|0\rangle^{\otimes N} + |1\rangle^{\otimes N}\right)$ as initial quantum resource, which can be depicted as follows.

Step 1: $P_1$ utilizes QRNG to generate the original random key $K_{1F}$, where $K_{1F} = \{K^1_{1F}, K^2_{1F}, \cdots, K^n_{1F}\}$, $K^r_{1F} \in \{0,1\}$, $r = 1,2,\cdots,n$. Similarly, $P_t$ also employs QRNG to generate the original random key $K_{tF}$, where $K_{tF} = \{K^1_{tF}, K^2_{tF}, \cdots, K^n_{tF}\}$, $K^r_{tF} \in \{0,1\}$, $t = 2,3,\cdots,N$, $r = 1,2,\cdots,n$. Then, $P_l$ calculates the hash value of her original random key $K_{lF}$ and publishes the result $h(K_{lF})$ to the other $N-1$ parties. Here, $h(K_{lF})$ represents the hash value of $K_{lF}$, and $l = 1,2,\cdots,N$.

Step 2: $P_1$ prepares $2^N n$ $N$-particle GHZ entangled states all in the state of $\left|\Psi^+_{0,0,\cdots,0}\right\rangle_N$. Then, $P_1$ divides these GHZ entangled states into $N$ sequences: $S_l = \{S^1_l, S^2_l, \cdots, S^{2^N n}_l\}$, where $S^j_l$ is the $j$th particle of $S_l$, $l = 1,2,\cdots,N$ and $j = 1,2,\cdots,2^N n$. Specifically, all $l$th particles of these GHZ entangled states form the ordered sequence $S_l$. $P_1$ sends the particles in $S_t$ to $P_t$ one by one, where $S_t = \{S^1_t, S^2_t, \cdots, S^{2^N n}_t\}$, $t = 2,3,\cdots,N$. Note that after $P_1$ sends the first particle to $P_t$, she sends a particle only after receiving the previous one.

Step 3: For the $j$th ($j = 1,2,\cdots,2^N n$) received particle in $S_t$ ($t = 2,3,\cdots,N$), $P_t$ randomly executes one of the following two operations: ① the CTRL operation, i.e., reflecting the $j$th received particle to $P_1$; and ② the SIFT operation, i.e., measuring the $j$th received particle with the $Z$ basis to obtain the bit value of the measurement result $M^j_t$, preparing a fresh particle in the same state as that she found and resending it to $P_1$. Note that $P_t$ chooses to SIFT and CTRL with equal probability; and there are only $2^{N-1} n$ particles $P_t$ chooses to SIFT. $P_t$ writes down $M_t = \{M^1_t, M^2_t, \cdots, M^{2^{N-1} n}_t\}$ when she chooses to SIFT.

Step 4: $P_t$ ($t = 2,3,\cdots,N$) tells $P_1$ the positions of $S_t$ where she chose to SIFT. $P_1$ performs different operations on the received particles according to $P_2, P_3, \cdots, P_N$'s choices, which can be summarized into Table 2. Case ① and Case ② are used to check whether the transmissions of $S_2, S_3, \cdots, S_N$ are secure or not; and Case ③ is used for not only checking the transmission security of $S_2, S_3, \cdots, S_N$ but also key agreement. Note that there are $2n$ positions where Case ③ happens; $P_1$ randomly choose $n$ positions among these $2n$ ones to check the transmission security of $S_2, S_3, \cdots, S_N$;

In Case ①, for the position where all of $P_2, P_3, \cdots, P_N$ chose to CTRL, $P_1$ performs Action 1. If there is no Eve on line, $P_1$'s measurement result should be $\left| \Psi^+_{0,0,0,\cdots,0} \right\rangle_N$;

In Case ②, for the position where not all of $P_2, P_3, \cdots, P_N$ chose the same operations, $P_1$ performs Action 2. If there is no Eve on line, the measurement results on the original particles from the parties who chose to SIFT, $P_1$'s measurement result on the corresponding particle in her own hand and $P_1$'s measurement results on the particles from the parties who chose to CTRL should be same. In order to detect whether an Eve exists or not, the parties who chose to SIFT need to tell $P_1$ their measurement results on the original particles;

In Case ③, for the position where all of $P_2, P_3, \cdots, P_N$ chose to SIFT, $P_1$ performs Action 3. If there is no Eve on line, $P_2, P_3, \cdots, P_N$'s measurement results on their original particles and $P_1$'s measurement result on the corresponding particle in her own hand should be same; and $P_1$'s measurement result on the particle from $P_t$ should be same to the fresh state generated by $P_t$. In order to detect whether an Eve exists or not, $P_t$ needs to tell $P_1$ her measurement results for the $n$ chosen positions;

If either of the error rates of these three Cases is abnormally high, the communication will be terminated immediately; otherwise, the communication will be continued.

Table 2  $P_1$'s actions corresponding to those of $P_2, P_3, \cdots, P_N$

| Case | $P_2, P_3, \cdots, P_N$ | $P_1$ |
|---|---|---|
| ① | All of them chose to CTRL | Action 1 |
| ② | Not all of them chose the same operations | Action 2 |
| ③ | All of them chose to SIFT | Action 3 |

Action 1: $P_1$ measures the particles from $P_2, P_3, \cdots, P_N$ together with the corresponding particle in her own hand with the $N$-particle GHZ basis;

Action 2: $P_1$ measures the particles from the parties who chose to CTRL and the corresponding particle in her own hand with the Z basis;

Action 3: $P_1$ measures the particles from $P_2, P_3, \cdots, P_N$ and the corresponding particle in her own hand with the Z basis.

Step 5: As for Case ③, after the $n$ particles used for security check are dropped out, the remaining $n$ ones are used for key agreement, which are named as INFO particles for simplicity. The bit values of $P_1$'s measurement results on the INFO particles in her own hand are denoted as

$M_{1F} = \{M_{1F}^1, M_{1F}^2, \cdots, M_{1F}^n\}$. $M_{tF} = \{M_{tF}^1, M_{tF}^2, \cdots, M_{tF}^n\}$ are the bits of $M_t$ corresponding to INFO particles, where $t = 2, 3, \cdots, N$. It is apparent that

$$M_{1F} = M_{2F} = \cdots = M_{NF}. \tag{24}$$

$P_l$ calculates

$$Q_{lF}^r = M_{lF}^r \oplus K_{lF}^r, \tag{25}$$

where $l = 1, 2, \cdots, N$, $r = 1, 2, \cdots, n$, and publishes $Q_{lF}$ to the other $N-1$ parties, where $Q_{lF} = \{Q_{lF}^1, Q_{lF}^2, \cdots, Q_{lF}^n\}$.

Step 6: According to $M_{lF}$, $P_l$ extracts $K_{vF}$ from $Q_{vF}$, where $l = 1, 2, \cdots, N$, $v = 1, 2, \cdots, N$ and $v \neq l$. Then, $P_l$ calculates the hash value of $K_{vF}$ to obtain the result $h'(K_{vF})$. If $h'(K_{vF}) = h(K_{vF})$ for $v = 1, 2, \cdots, N$ and $v \neq l$, $P_l$ will accept $K_F = K_{1F} \oplus K_{2F} \oplus \cdots \oplus K_{NF}$ as the final shared key. If any of $P_1, P_2, \cdots, P_N$ refuses $K_F = K_{1F} \oplus K_{2F} \oplus \cdots \oplus K_{NF}$ as the final shared key, the protocol will be terminated and restarted from the beginning.

## 5  Discussions and conclusions

In a quantum communication protocol, we usually use the qubit efficiency to evaluate its performance of efficiency, which is defined as [4]

$$\eta = \frac{f}{q+c}, \tag{26}$$

where $f$ is the number of bits of the final shared private key, $q$ represents the number of consumed qubits, and $c$ denotes the number of classical bits needed for the classical communication. Note that the classical resources necessary for eavesdropping check are ignored here. In our $N$-party SQKA protocol, $P_1, P_2, \cdots, P_N$ can equally establish a $n$-bit final shared private key, so it has $f = n$. $P_1$ needs to prepare $2^N n$ $N$-particle GHZ entangled states and send the particles in $S_t$ to $P_t$ one by one; and when $P_t$ chooses to SIFT the received particles in $S_t$, she needs to generate $2^{N-1} n$ new particles, where $t = 2, 3, \cdots, N$. As a result, it has $q = 2^N n \times N + 2^{N-1} n \times (N-1) = 2^{N-1} n (3N-1)$. $P_l$ needs to publish $h(K_{lF})$ and $Q_{lF}$ to the other $N-1$ parties, where $l = 1, 2, \cdots, N$. Hence, it has $c = mN + nN$, where $m$ is the length of $h(K_{lF})$. Therefore, the qubit efficiency of our $N$-party SQKA protocol is $\eta = \dfrac{n}{2^{N-1} n (3N-1) + mN + nN}$.

To sum up, in this paper, in order to realize the goal that a quantum party and two classical parties equally contribute to the generation of a shared private key over quantum channels, we design a novel single-state three-party SQKA protocol, which only uses one kind of three-particle GHZ entangled states as initial quantum resource. We validate its security in detail, and show that it can resist both the outside attack and the participant attack. The proposed single-state three-party SQKA protocol has several good features: (1) it only adopts one kind of three-particle GHZ entangled states as initial quantum resource; (2) it doesn't need pre-shared keys among different parties; (3) it doesn't need unitary operations or quantum entanglement swapping. Finally, we generalize it into the case of $N$-party by only employing one kind of $N$-particle GHZ entangled states as initial quantum resource, which inherits the nice features of its three-party counterpart.

**Acknowledgments**

The authors would like to thank the anonymous reviewers for their valuable comments that help enhancing the quality of this paper. Funding by the National Natural Science Foundation of China (Grant No.62071430 and No.61871347), the Fundamental Research Funds for the Provincial Universities of Zhejiang (Grant No.JRK21002) and Zhejiang Gongshang University, Zhejiang Provincial Key Laboratory of New Network Standards and Technologies (No. 2013E10012) is gratefully acknowledged.